\documentclass[aps,twocolumn,showpacs,preprintnumbers,amsmath,amssymb,superscriptaddress,pra,nofootinbib]{revtex4-1}

\usepackage{graphicx}
\usepackage{dcolumn}
\usepackage{bm}
\usepackage{graphics}
\usepackage{subfigure}
\usepackage{graphicx}
\usepackage{epsfig}
\usepackage{epstopdf}
\usepackage{color}
\usepackage{mathtools}
\usepackage{amssymb,amsmath}
\usepackage{graphicx}
\usepackage{bookmark}
\usepackage{lmodern}
\usepackage{textcomp}
\usepackage{longtable}
\usepackage{braket}
\usepackage{pdfpages}
\usepackage{epstopdf}
\usepackage{bbm}
\usepackage{mathtools}

\newcommand{\eps}{{\varepsilon}}
\newcommand{\vk}{{\bf k}}
\newcommand{\vp}{{\bf p}}
\newcommand{\vq}{{\bf q}}
\newcommand{\vP}{{\bf P}}

\begin{document}

\title{Efimov correlations in strongly interacting Bose gases}

\author{Marcus Barth}
\email{marcus.barth@ph.tum.de}
\affiliation{Technische\,Universit{\"a}t\,M{\"u}nchen,\,Physik\,Department,\,
James-Franck-Strasse,\,85748 Garching,\,Germany}

\author{Johannes Hofmann}
\email{hofmann@umd.edu}
\affiliation{Condensed Matter Theory Center and Joint Quantum Institute, Department of Physics, University of Maryland, College Park, Maryland 20742-4111 USA}

\date{\today}

\begin{abstract}
We compute the virial coefficients, the contact parameters, and the momentum distribution of a strongly interacting three-dimensional Bose gas by means of a virial expansion up to third order in the fugacity, which takes into account three-body correlations exactly. Our results characterize the nondegenerate regime of the interacting Bose gas, where the thermal wavelength is smaller than the interparticle spacing but the scattering length may be arbitrarily large. We observe a rapid variation of the third virial coefficient as the scattering length is tuned across the three-atom and the atom-dimer thresholds. The momentum distribution at unitarity displays a universal high-momentum tail with a log-periodic momentum dependence, which is a direct signature of Efimov physics. We provide a quantitative description of the momentum distribution at high momentum as measured by P. Makotyn {\it et al.} [Nat. Phys. {\bf 10}, 116 (2014)], and our calculations indicate that the lowest trimer state might not be occupied 
in 
the experiment. Our results allow for a spectroscopy of Efimov states in the unitary limit.
\end{abstract}


\maketitle

\section{Introduction}

The classical problem in Newtonian mechanics of finding analytical stable orbits of three planets has only a few known solutions which, moreover, exist only for certain special conditions~\cite{Suvakov13}. Curiously, a generic analog of this problem in quantum mechanics, three neutral bosons of mass $m$ with a universal short-range interaction, was solved analytically by Efimov four decades ago~\cite{efimov70,efimov71,braaten06}. If the effective range of the interparticle interaction (set by the van der Waals length $\ell_{vdW}$) is small compared to other length scales, the two-body interaction is solely characterized by the scattering length $a$. In the unitary limit of infinite scattering length, Efimov found an infinite number of three-particle bound states with energy $E_T^{(j)}=\tfrac{\kappa_*^2}{m}(e^{\pi/s_0})^{-2j}$, where $\kappa_*$ is a universal three-body parameter, $j$ is an integer, and $e^{\pi/s_0}\approx22.7$. Originally  predicted in  nuclear physics, Efimov states were observed in atom-
loss experiments and radio-frequency spectroscopy measurements in Bose gases~\cite{Kraemer06,Knoop09,zaccanti09,gross09,pollack09}, and three-component Fermi gases~\cite{Ottenstein08,Huckans09,Williams09,Lompe10,Nakajima11}, as well as mass-imbalanced mixtures~\cite{Barotini09,Tung14,Pires14}. Recently, the Efimov trimer of ${}^4$He has also been observed experimentally~\cite{Kunitski15}. Efimov physics is thus a very general phenomenon, and in addition to these many experimental realizations, Efimov states are also predicted to exist, for example, in $p$-wave interacting quantum gases~\cite{Nishida13b}, in condensed-matter quantum magnets~\cite{Nishida13a}, and in mass-imbalanced two-component Fermi gases~\cite{Efimov73,petrov03}. Most theoretical work focuses on few-body aspects of Efimov physics and experiments are commonly explained using few-body theory~\cite{braaten06}. In this paper, by contrast, we study the interacting Bose gas at finite density and temperature and establish signatures of three-body 
correlations in this system. The aim of this work is to provide a first exact quantitative calculation of three-body effects on a many-body system, which is relevant from a fundamental point of view and can be compared with current experimental work. The main result of this paper is a calculation of the full momentum distribution in the nondegenerate limit for arbitrary values of scattering length and three-body parameter $\kappa_*$.

Most experiments on Bose gases in equilibrium are restricted to the weak-interaction regime with few exceptions~\cite{Navon11,ha13}.  This is due to enhanced three-body losses at finite density with increasing interaction strength which deplete the gas with a rate $\dot{N}=L_3 n^2 N$~\cite{braaten06}. At zero temperature, the loss coefficient scales as $L_3(T=0) \sim a^4$~\cite{fedichev96,esben99,esry99} and saturates to $L_3(T)\sim\lambda_T^4\sim1/T^2$~\cite{dincao04} in the unitary limit at high temperature ($\lambda_T=\sqrt{2\pi/mT}$ denotes the thermal wavelength, and we set $\hbar=k_B=1$). In a series of recent hallmark experiments, the loss rate of a strongly interacting Bose gas at finite temperature was measured~\cite{fletcher13,rem13,makotyn14,Eismann15}. It turns out that in the nondegenerate limit $n\lambda_T^3\ll1$, the two-body scattering rate $\gamma_2=n\sigma v\sim n\lambda_T$ is much larger compared to the three-body loss rate $\gamma_3=L_3n^2\sim n^2\lambda_T^4$, so that the gas can reach an 
equilibrium state before a significant fraction of particles is lost. Indeed, this is corroborated by a recent experiment that quenches a weakly interacting Bose-Einstein condensate to the unitary limit and measures the momentum distribution, which approaches a stationary equilibrium distribution shortly after the quench~\cite{makotyn14}. In the nondegenerate regime, we furthermore expect that the system is thermodynamically stable~\cite{Stoof94,Mueller00,Jeon02,Li12,Pitaecki14,vanDijk14}. The experiments~\cite{fletcher13,rem13,makotyn14,Eismann15} demonstrate that strongly interacting Bose gases, for which three-body correlations turn out to be essential, can be experimentally prepared and studied.

Given this experimental progress, there is a need for an accurate quantitative description of a strongly interacting Bose gas which includes three-body effects. Yet, typical many-body methods, such as the $T$-matrix approximation, do not take into account three-body correlations, and other methods must be developed. As discussed above, strongly interacting Bose gases can reach an equilibrium state in the nondegenerate limit where the thermal wavelength is smaller compared to the interparticle spacing. This regime should be accurately described by a virial expansion, and in this paper, we develop the virial expansion for the strongly interacting Bose gas including three-body corrections.

This paper is structured as follows: in Sec.~\ref{sec:methods}, we discuss the virial expansion and introduce the field-theoretical model for cold bosons that we shall use throughout the paper. Following this, we outline how three-body effects of the Bose gas are taken into account in the virial expansion in a diagrammatic framework. Section~\ref{sec:results} presents the results of our work: we begin in Sec.~\ref{sec:virialcoefficients} by discussing our results for the first three virial coefficients. The second virial coefficient depends only on the scattering length and interpolates between a scattering-dominated limit for negative scattering length and a dimer-dominated limit for positive scattering length. The third virial coefficient depends not only on the scattering length but also sensitively on the three-body parameter $\kappa_*$. In particular, we find a rapid variation in the virial coefficient as the scattering length is tuned across the three-body thresholds where the system can support a 
single trimer bound state. In Sec.~\ref{sec:contacts}, we provide results for the two-body and three-body contact parameters, which describe various thermodynamic properties of the Bose gas and set the magnitude of the asymptotic form of many response functions. Beside these thermodynamic quantities, we also develop the virial expansion for the full momentum distribution including three-body correlations, and the calculation of the momentum distribution, which constitutes the main result of our work, is presented in Sec.~\ref{sec:momentum}. We discuss in detail the behavior of the momentum distribution as a function of both the scattering length and the three-body parameter. Our results are in excellent agreement with universal relations that govern the high-momentum tail of the momentum distribution, where we determine the contact parameters from the independent calculation presented in Sec.~\ref{sec:contacts}. Especially, we find very good agreement with a subleading log-periodic high-momentum tail which 
is a direct manifestation of Efimov physics. We also briefly make a comparison with a calculation of the Fermi momentum distribution, for which the virial expansion provides only a quantitative correction but does not show new qualitative three-body effects. Our results for the bosonic momentum distribution can be directly compared with the recent experiment by Makotyn {\it et al.}~\cite{makotyn14}, which is in the nondegenerate regime following a quench to the unitary limit. We provide a fit to the experimental data which contains only the temperature as a single fitting parameter. Our calculation provides an accurate quantitative description of the experimental results. Surprisingly, the best fit is found when the lowest Efimov trimer state is excluded from the calculation, indicating that the lowest trimer state is not occupied in the experiment~\cite{makotyn14}. The paper is concluded with a summary and conclusion in Sec.~\ref{sec:summary}. There are two appendices which contain details of the 
calculation: Appendix~\ref{sec:appA} describes the solution of the three-body scattering amplitude, and Appendix~\ref{sec:appB} contains intermediate results for diagrams that contribute to the virial expansion of the Bose gas.

\begin{figure}[t!]
\begin{center}
\subfigure[]{\raisebox{0.0cm}{\scalebox{0.24}{\includegraphics{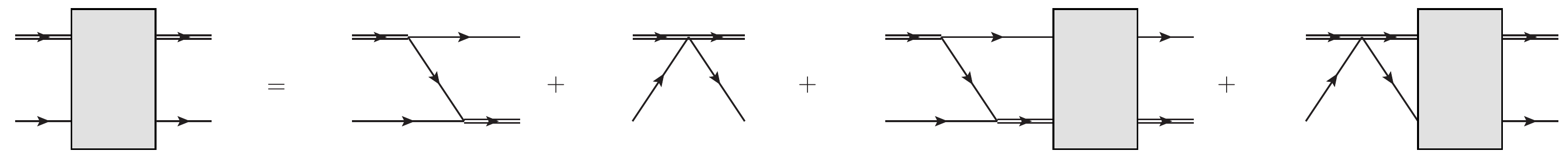}}}\label{fig:3bTmatrix}}\\
\subfigure[]{\raisebox{0.0cm}{\scalebox{0.3}{\includegraphics{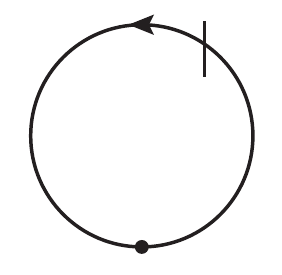}}}\label{fig:Da}}
\subfigure[]{\raisebox{0.0cm}{\scalebox{0.3}{\includegraphics{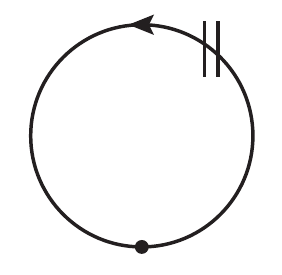}}}\label{fig:Db}}
\subfigure[]{\raisebox{0.0cm}{\scalebox{0.3}{\includegraphics{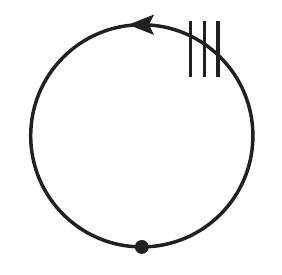}}}\label{fig:Dc}}
\subfigure[]{\raisebox{0.0cm}{\scalebox{0.3}{\includegraphics{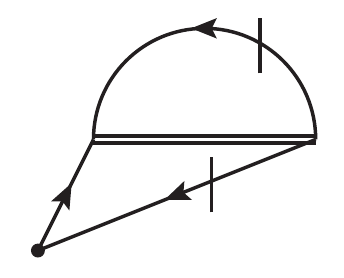}}}\label{fig:Dd}}
\subfigure[]{\raisebox{0.0cm}{\scalebox{0.3}{\includegraphics{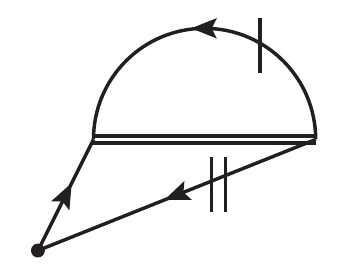}}}\label{fig:De}}
\subfigure[]{\raisebox{0.0cm}{\scalebox{0.3}{\includegraphics{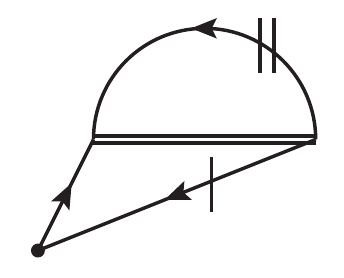}}}\label{fig:Df}}
\subfigure[]{\raisebox{0.0cm}{\scalebox{0.3}{\includegraphics{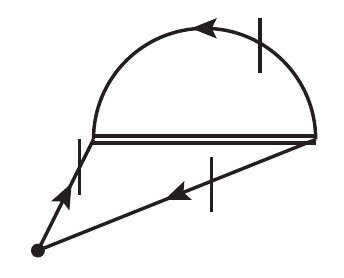}}}\label{fig:Dg}}\\
\subfigure[]{\raisebox{0.0cm}{\scalebox{0.2}{\includegraphics{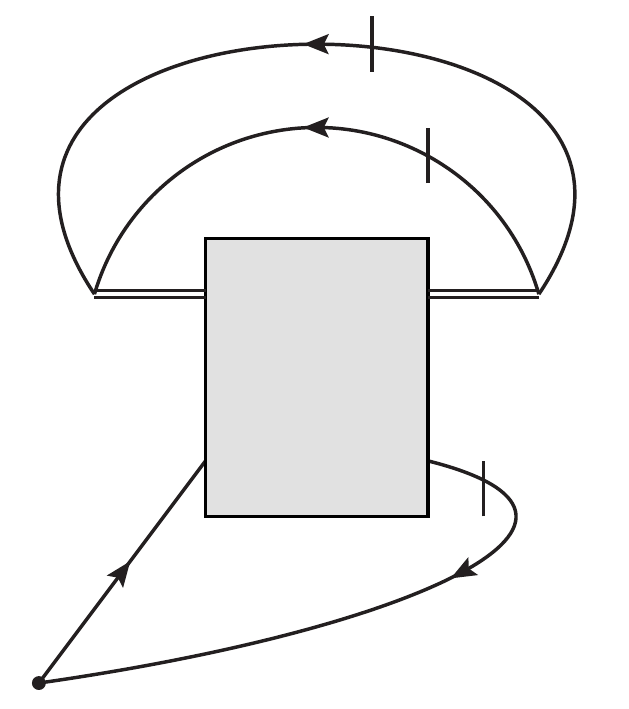}}}\label{fig:Dh}}
\subfigure[]{\raisebox{0.0cm}{\scalebox{0.2}{\includegraphics{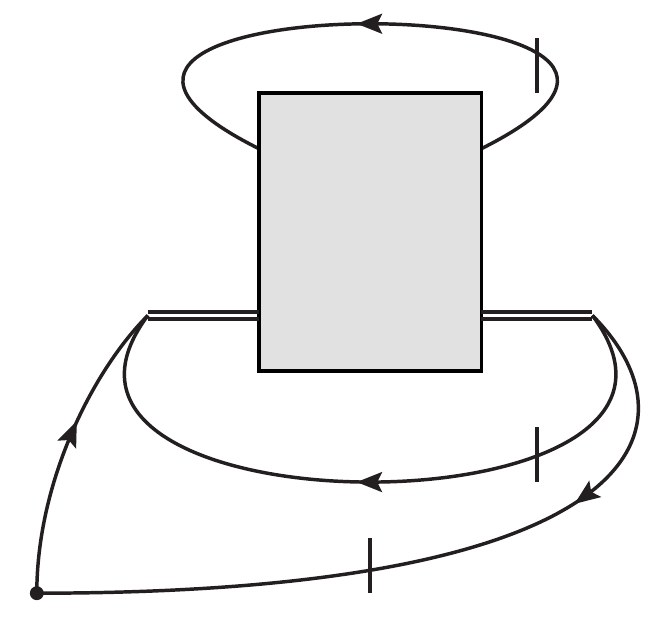}}}\label{fig:Di}}
\subfigure[]{\raisebox{0.0cm}{\scalebox{0.2}{\includegraphics{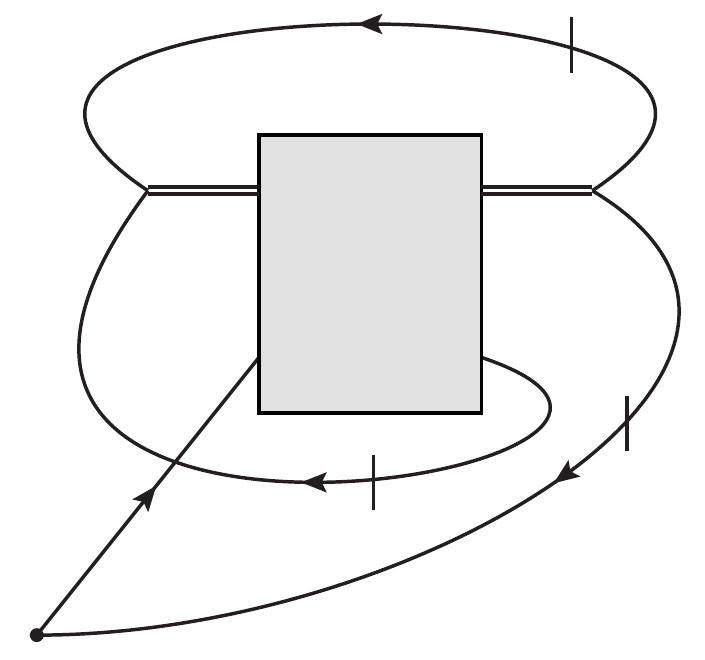}}}\label{fig:Dj}}
\subfigure[]{\raisebox{0.0cm}{\scalebox{0.2}{\includegraphics{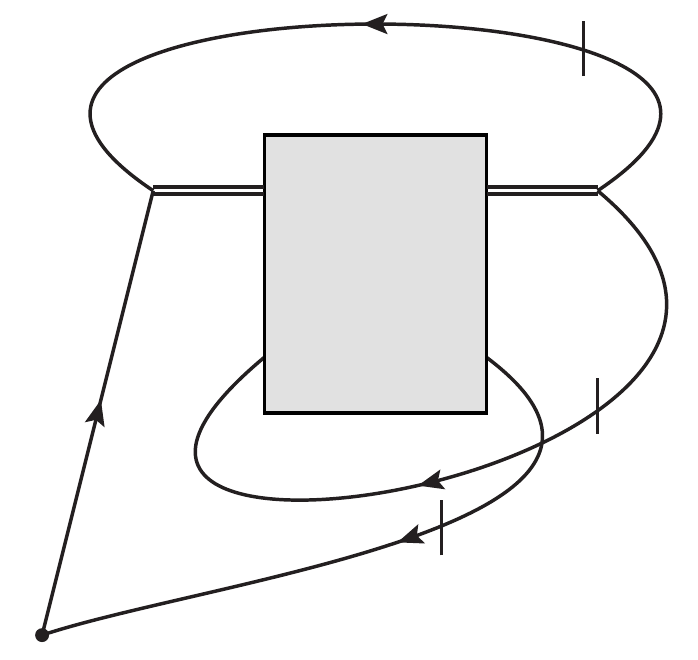}}}\label{fig:Dk}}
\subfigure[]{\raisebox{0.0cm}{\scalebox{0.2}{\includegraphics{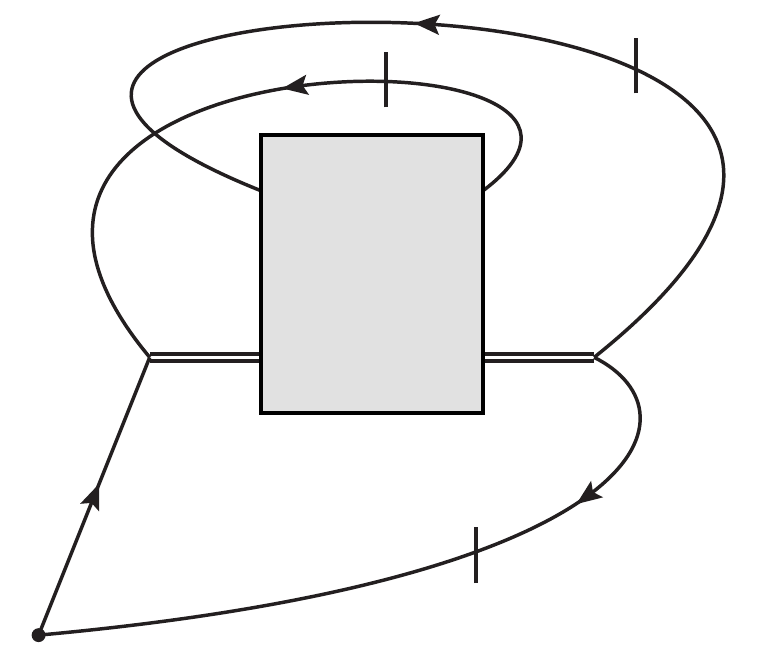}}}\label{fig:Dl}}
\caption{(a) Integral equation for the three-body scattering matrix, denoted by a gray box. Simple lines denote atom propagators, and double lines are dimer propagators. (b)-(m) Diagrams that contribute to the density and the momentum distribution up to third order in the fugacity.}
\label{fig:diagrams}
\end{center}
\end{figure}

\section{Methods}\label{sec:methods}

In this paper, we characterize the strongly interacting Bose gas in the normal phase by performing a virial expansion, allowing us to link few-body physics to the properties of an interacting many-body system in a systematic way. In particular, in addition to the thermodynamic virial coefficients, we compute the full momentum distribution and establish Efimov signatures in this correlation function. The virial expansion is valid if the thermal wavelength is much smaller than the interparticle spacing ($n\lambda_T^3\ll1$) and is therefore ideally suited to describe the current experimental work~\cite{fletcher13,rem13,makotyn14,Eismann15}. Any operator can be expressed as a sum of separate traces over connected $N$-particle sectors:
\begin{equation}
\langle {\cal O} \rangle = \frac{{\rm Tr} \, {\cal O} e^{-\beta (H-\mu N)}}{{\rm Tr} e^{-\beta (H-\mu N)}} = \sum_{N=0}^\infty z^N {\rm tr}_{N} {\cal O} e^{-\beta H} . \label{eq:defvirial}
\end{equation}
In the nondegenerate regime $n\lambda_T^3\ll1$, the fugacity $z=e^{\beta \mu}$ is a small parameter and the expansion in Eq.~\eqref{eq:defvirial} can be truncated after the first few terms. By performing the expansion up to third order in $z$, we fully include three-body correlations. Previous early work on the virial coefficients of Bose gases is~\cite{Pais59,Dashen69} and more recent applications are~\cite{Bedaque03,castin13,laurent14}. A previous numerical study of the virial coefficients was carried out by Bedaque and Rupak~\cite{Bedaque03}, but they have not included all the relevant diagrams (see Fig. \ref{fig:diagrams}). Note that the virial coefficients at unitarity can even be determined analytically~\cite{castin13}. The virial expansion has also been successfully applied to Fermi gases; see Ref.~\cite{Liu13} for a review. In the following, we provide numerical results for the first three virial coefficients (both at unitarity and at finite  scattering length) as well as the two-body and the three-
body contacts. An essential new point of our analysis is that we develop the virial expansion for the full Green's function, which allows us to compute the momentum distribution. The momentum distribution exhibits a universal high-momentum tail that shows direct three-body Efimov correlations. We compare our calculations with the recent experiment~\cite{makotyn14}.

The Lagrangian density of an interacting Bose gas with large scattering length takes the form~\cite{Bedaque99a,braaten06}
\begin{eqnarray}
{\cal L} &=& \phi^\dagger \biggl(i \partial_t + \frac{\nabla^2}{2m}\biggr) \phi +  \frac{g_2}{4} d^\dagger d^{}  - \frac{g_2}{4} \biggl(d^\dagger \phi \phi + \phi^\dagger \phi^\dagger d\biggr) \nonumber \\
&-& \frac{g_3}{36} \phi^\dagger d^\dagger d \phi , \label{eq:lagrangian}
\end{eqnarray}
where $\phi^\dagger$ creates a boson and $d$ is an auxiliary dimer field. The bare coupling constants $g_2$ and $g_3$ can be written as $\tfrac{1}{g_2}=\tfrac{m}{8\pi a}-\tfrac{m\Lambda}{4 \pi^2}$ and $g_3=-9mg_2^2\tfrac{H(\Lambda)}{\Lambda^2}$ with $H(\Lambda)\sim\tfrac{\cos(s_0\ln \Lambda/\Lambda_*+\arctan s_0)}{\cos(s_0\ln\Lambda/\Lambda_*-\arctan s_0)}$, where $\Lambda$ is a cutoff regulator and $\Lambda_*$ is a renormalized three-body parameter~\cite{Bedaque99a,Bedaque99b,braaten06}. For positive scattering length, there is a dimer bound state with energy $E_D=\tfrac{1}{m a^2}$. In addition, theory~\eqref{eq:lagrangian} has an infinite number of arbitrarily deep Efimov bound states. We avoid the Thomas collapse~\cite{thomas35} of the zero range model by specifying a lowest Efimov trimer state, the energy of which we denote by $E_T$. The parameter $\Lambda_*$ is related to the wavenumber of the lowest Efimov trimer at unitarity via $\kappa_*=0.38\Lambda_*$. Each trimer branch hits the continuum of 
scattering states at the three-body threshold scattering length $a_-$ and terminates at positive scattering length at the atom-dimer threshold $a_*$. Our calculation is based on a diagrammatic representation developed for the virial coefficients of a Fermi gas~\cite{leyronas11}, which we generalize to the Bose case. Crucially, we show that this method can be applied to any correlator and we identify the three-body effects on the momentum distribution. The starting point is an expansion in imaginary time, where it turns out that the full dependence on the fugacity $z$ is encoded in the free propagator, which can be expanded in powers of $z$. In Fig.~\ref{fig:diagrams}, we denote the $j$th-order contribution to the propagator by a line that is slashed $j$ times. The zero-order contribution can only propagate forward in imaginary time, and backward-propagating lines contribute higher powers of $z$. This provides a diagrammatic representation for the virial expansion of any correlation function. The Feynman 
diagrams constructed in this way contain subdiagrams that involve repeated scattering of either two or three particles. These are the scattering matrices of the few-body problem, and we denote them by a double line and a box, respectively.

\section{Results}\label{sec:results}

\subsection{Virial coefficients}\label{sec:virialcoefficients}
\begin{figure}[t!]
\begin{center}
\scalebox{0.95}{\includegraphics[width=\columnwidth]{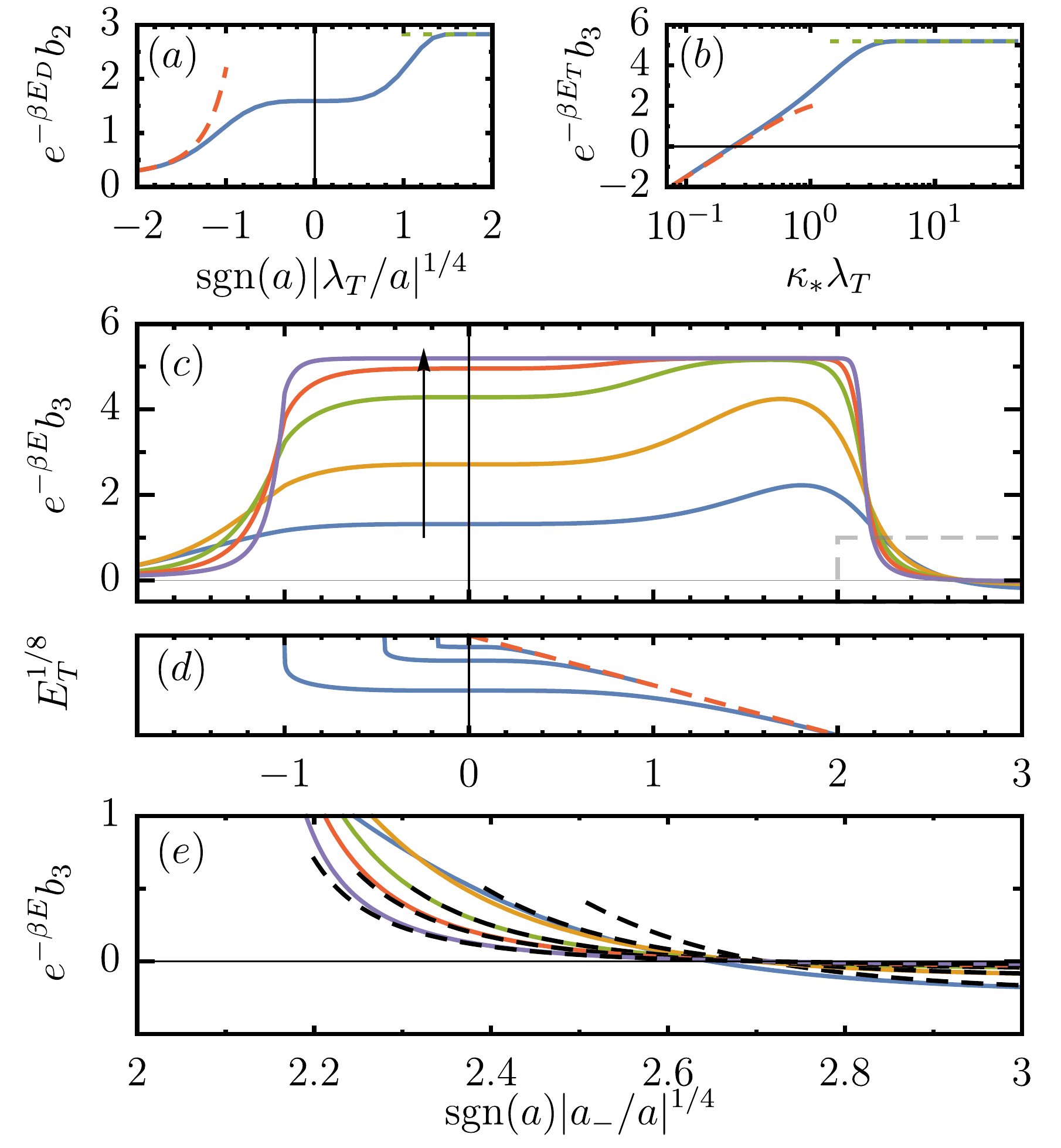}}
\caption{
(Color online) (a) Second virial coefficient as a function of scattering length. (b) Third virial coefficient at unitarity as a function of the three-body parameter. The dashed red and dotted green lines in (a) and (b) denote the analytical asymptotic results~\cite{Bedaque03,castin13}. (c) Third virial coefficient as a function of scattering length for fixed (along the arrow) $\kappa_*\lambda_T=0.5,1,2,3,5$. (d) Bound-state spectrum. Blue solid lines denote trimer branches, and the red dashed line is the dimer bound-state energy. (e) Third virial coefficient at small and positive scattering length $a \ll \lambda_T$. Dashed lines denote the asymptotic result based on the atom-dimer Beth-Uhlenbeck formula~\eqref{eq:bethuhlenbeck}.
}
\label{fig:virial}
\end{center}
\end{figure}

\begin{figure}[t!]
\begin{center}
\scalebox{0.95}{
\includegraphics{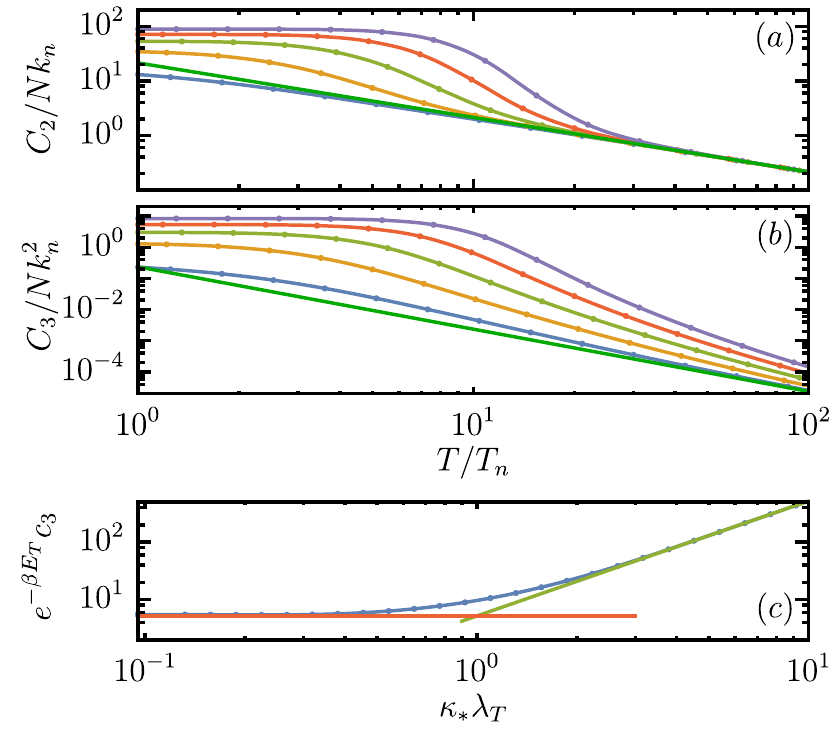}
}
\caption{(Color online) (a) Two-body contact as a function of temperature. Three-body contact at unitarity as a function of (b) temperature and (c) the three-body parameter. The lines in~(a) and (b) correspond to (from bottom to top) $\kappa_*/k_n=1,2,3,4,$ and $5$. The green lines in (a) and (b) denote the asymptotic high-temperature results. The red and green lines in (c) denote the asymptotic results as stated in the main text.}
\label{fig:contact}
\end{center}
\end{figure}

We begin by computing the virial expansion of the density, which is related to the virial coefficients via~\cite{Pais59}
\begin{equation}
\lambda_T^3 n = b_1 z + 2 b_2 z^2 + 3 b_3 z^3 + {\it O}(z^4) .
\end{equation}
The diagrams contributing to the density up to ${\it O}(z^3)$ are shown in Fig.~\ref{fig:diagrams}, with Fig.~\ref{fig:Da} contributing to $b_1$, Figs.~\ref{fig:Db} and~\ref{fig:Dd} contributing to $b_2$, and all other diagrams contributing to the third virial coefficient $b_3$. Figures ~\ref{fig:Da}-\ref{fig:Dc} give the standard result $b_j^{(0)} = j^{-5/2} $ for the noninteracting Bose gas, while Fig.~\ref{fig:Dd} yields the Beth-Uhlenbeck interaction correction~\cite{Bedaque03,beth37}. Figures~\ref{fig:Dh}-\ref{fig:Dl} have to be evaluated numerically, and the three-body amplitude in the integrand is obtained by solving the Skorniakov--Ter-Martirosian integral equation~\cite{skorniakov57} shown in Fig.~\ref{fig:3bTmatrix}. Extensive details of our calculation are relegated to the appendices. Figure~\ref{fig:virial} shows our results for the second and third virial coefficients. Both the second [Fig.~\ref{fig:virial}(a)] and third virial coefficients [Fig.~\ref{fig:virial}(b)] agree with the analytical 
results~\cite{Bedaque03,castin13}, providing an independent check of our calculations. Figure~\ref{fig:virial}(c) shows the third virial coefficient as a function of scattering length for various values of $\kappa_*\lambda_T$. We rescale the scattering length by $a_-$ for each $\kappa_*$ and plot the reduced virial coefficient $e^{-\beta E}b_3$, where $E$ is the lowest bound-state energy ($E=E_T$ for $a^{ -1}<a_*^{ -1}$, $E=E_D$ for $a^{ -1}>a_*^{ -1}$, and $E=0$ if there is no bound state). In the scattering-dominated limit $\kappa_*\lambda_T\ll1$, the virial coefficient increases smoothly as the scattering length is tuned across the unitary limit and then decreases for $a^{ -1}>a_*^{-1}$, assuming negative values. In the trimer limit $\kappa_*\lambda_T\gg1$, the reduced virial coefficient is very small if no Efimov bound state is contained in the spectrum and rapidly jumps to the trimer-dominated result $e^{-\beta E_T} b_3 = 3\sqrt{3}$~\cite{castin13,Pais59} otherwise. Figure~\ref{fig:virial}(e) magnifies 
$b_3$ at small scattering length $a^{-1} > a_*^{-1}$ for which only the dimer bound state exists. In this limit, we find that the virial coefficient is well described by an effective two-body Beth-Uhlenbeck formula for the scattering of atoms and dimers~\cite{beth37,[{An analogous result holds for the fermionic case with $a_{\rm ad} = 1.18 a$, see }] ngampruetikorn15}
\begin{equation}
 e^{-\beta E_D} b_3  \approx - 3^{3/2}  \int_0^\infty \frac{dk}{\pi} e^{-\beta \frac{3 k^2}{4m}} \frac{a_{\rm ad}}{1+a_{\rm ad}^2 k^2}, \label{eq:bethuhlenbeck}
\end{equation}
where $a_{\rm ad}(a,\kappa_*)$ is the atom-dimer scattering length~\cite{braaten06}. Note that the zero-range description~\eqref{eq:lagrangian} holds if the scattering length and the thermal wavelength are large compared to the effective range: $a,\lambda_T>\ell_{vdW}$ (however, $a$ and $\lambda_T$ can have arbitrary ratios). Since the three-body parameter is found to be universal with $\kappa_*\approx 0.2/\ell_{vdW}$~\cite{schmidt12,wang12}, we expect that in the scattering-dominated limit, where $k_BT$ is larger than the energy of the lowest Efimov trimer, $\kappa_*\lambda_T\ll1$, there are effective range corrections to the results of this paper.

\begin{figure}[t!]
\begin{center}
\scalebox{0.95}{
\includegraphics{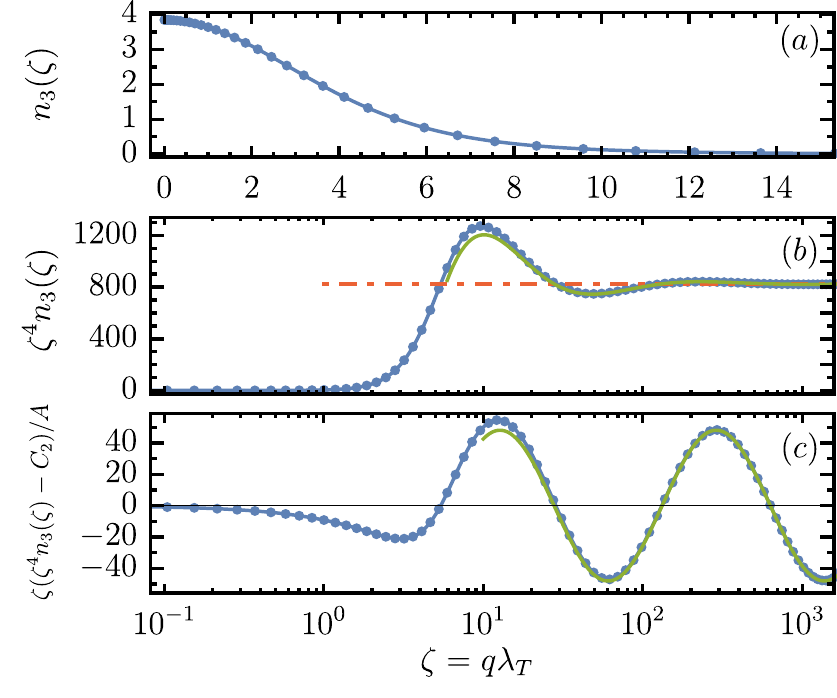}
}
\caption{(Color online) Third-order contribution $n_3(\zeta)$ to the momentum distribution at unitarity with $\kappa_*\lambda_T=3$ as a function of dimensionless momentum $\zeta=q\lambda_T$. Points denote numerical results; the solid blue line is a guide to the eye. The green and dot-dashed orange lines denote the asymptotic Tan relations~\eqref{eq:highmomentumtail} obtained with $e^{-\beta E_T}c_{2,3}=825.2$ and $e^{-\beta E_T}c_3=48.1$.}
\label{fig:momentum}
\end{center}
\end{figure}

\begin{figure*}[t!]
\begin{center}
\subfigure[\label{fig:momainv-a} $\lambda_T/a=-3.16$, $\beta E = 0$]{\scalebox{0.32}{\includegraphics{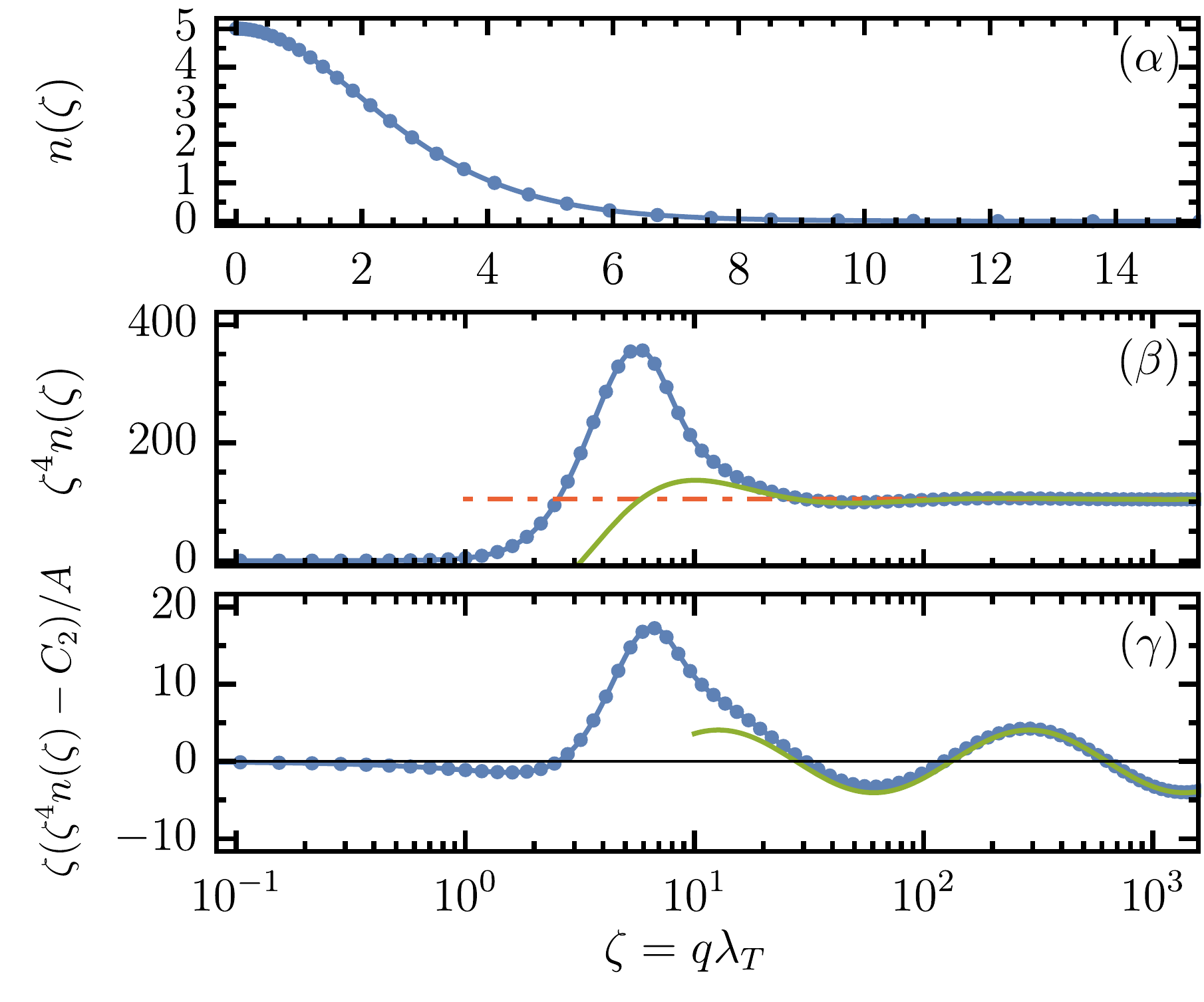}}}
\subfigure[\label{fig:momainv-b} $\lambda_T/a=0$, $\beta E = 1.43$]{\scalebox{0.32}{\includegraphics{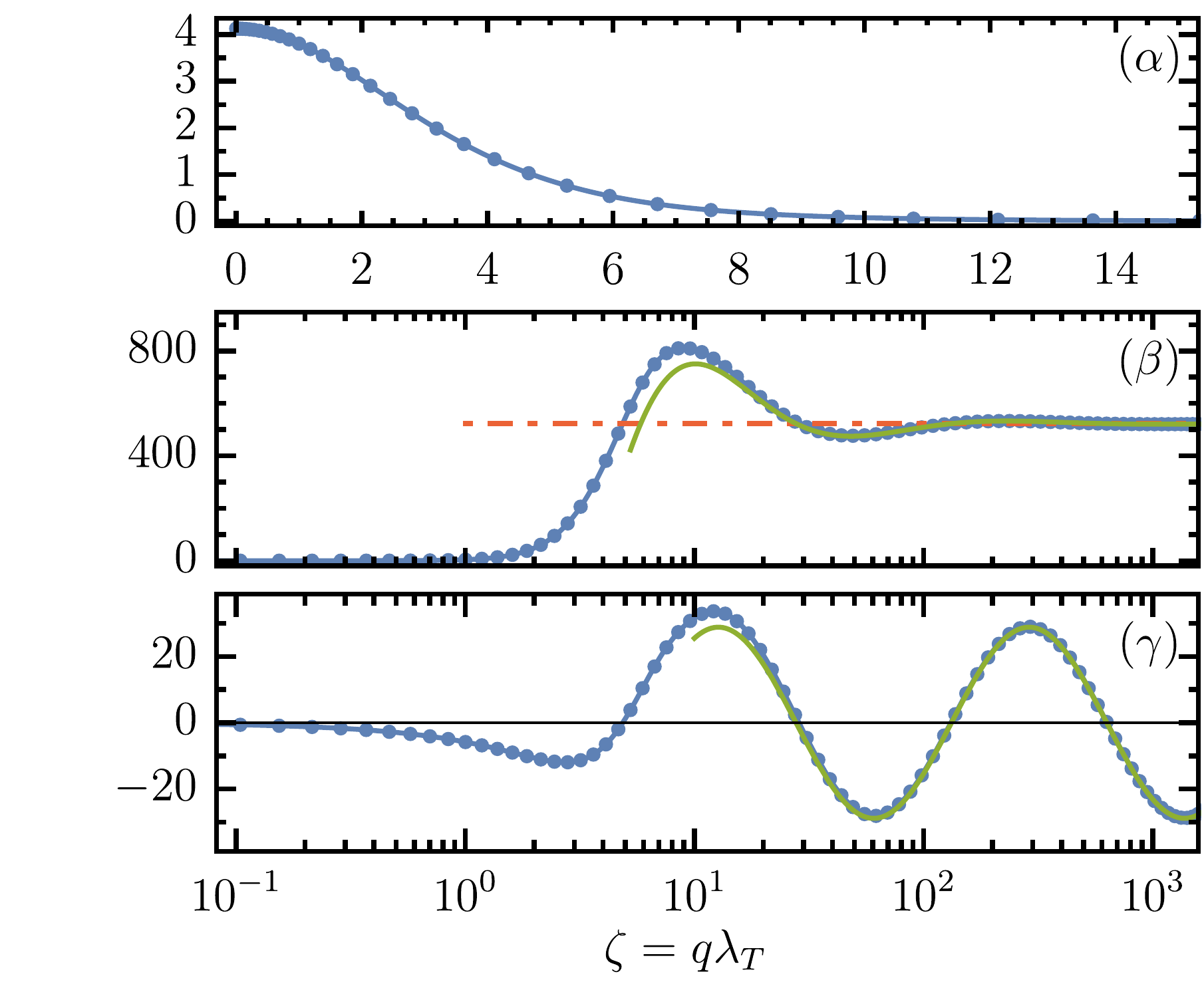}}}
\subfigure[\label{fig:momainv-c} $\lambda_T/a=3.16$, $\beta E = 5.84$]{\scalebox{0.32}{\includegraphics{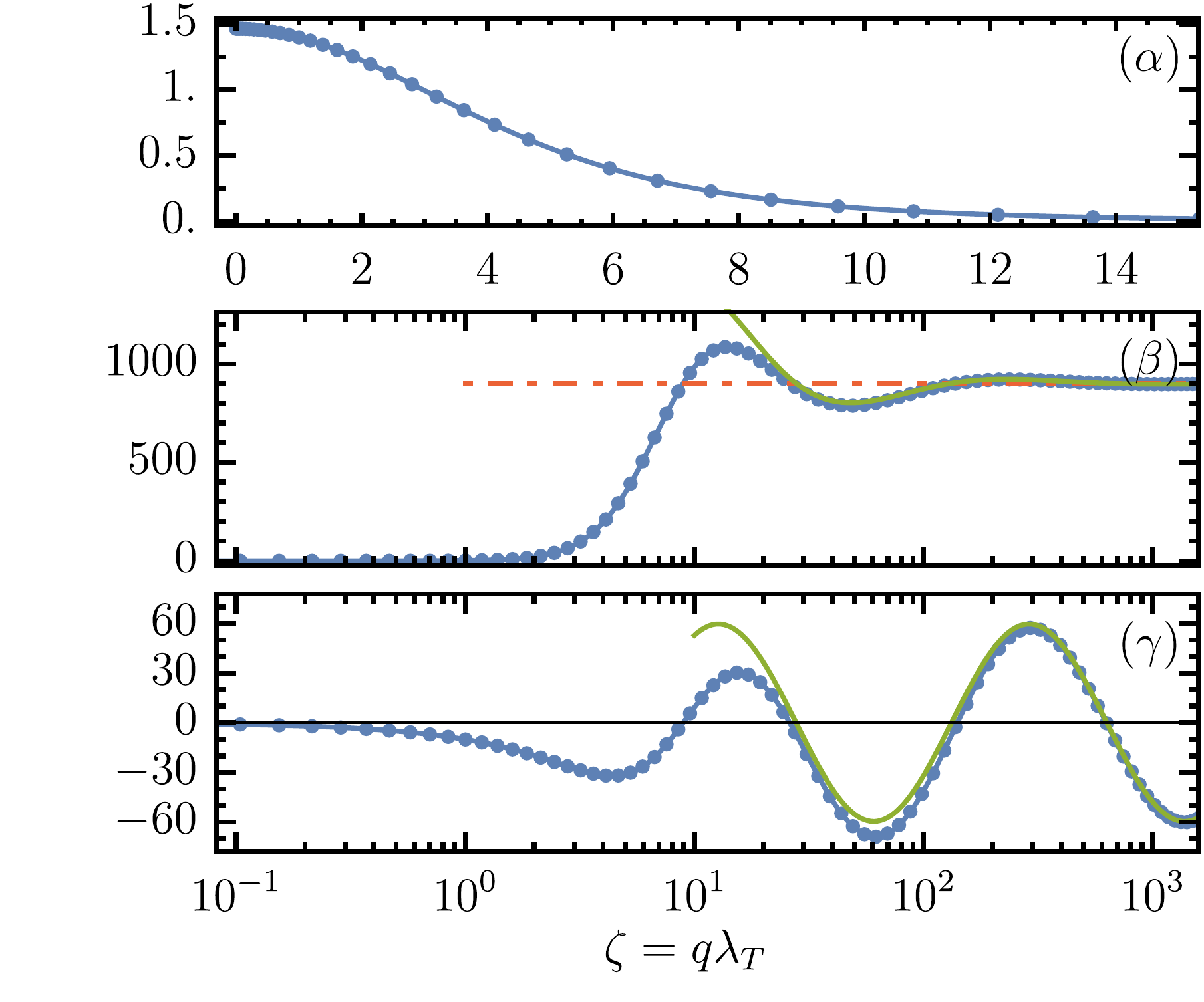}}}
\caption{(Color online) Full momentum distribution $n(\zeta)$ up to third order in the fugacity with $\kappa_*\lambda_T=3$ for three different scattering lengths $\lambda_T / a =-3.16,0$, and $3.16$. Fugacities have been chosen such that $z^3 e^{\beta E} = 0.6$. The notation is the same as in Fig.~\ref{fig:momentum}.}
\label{fig:momentumainv}
\end{center}
\end{figure*}
\begin{figure*}[t!]
\begin{center}
\subfigure[\ $\kappa_*\lambda_T=1$, $\beta E = 0.16$]{\scalebox{0.32}{\includegraphics{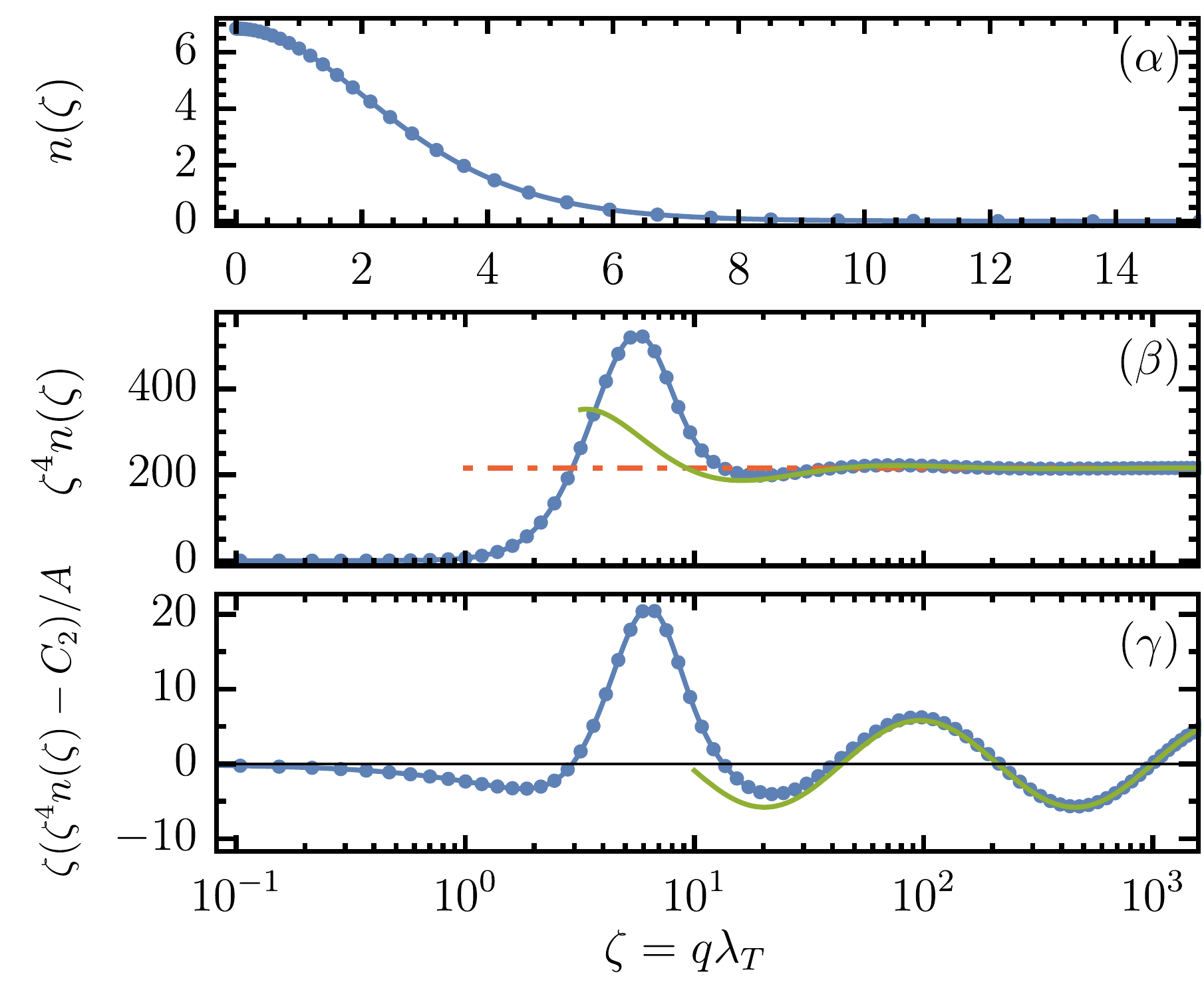}}}
\subfigure[\ $\kappa_*\lambda_T=3$, $\beta E = 1.43$]{\scalebox{0.32}{\includegraphics{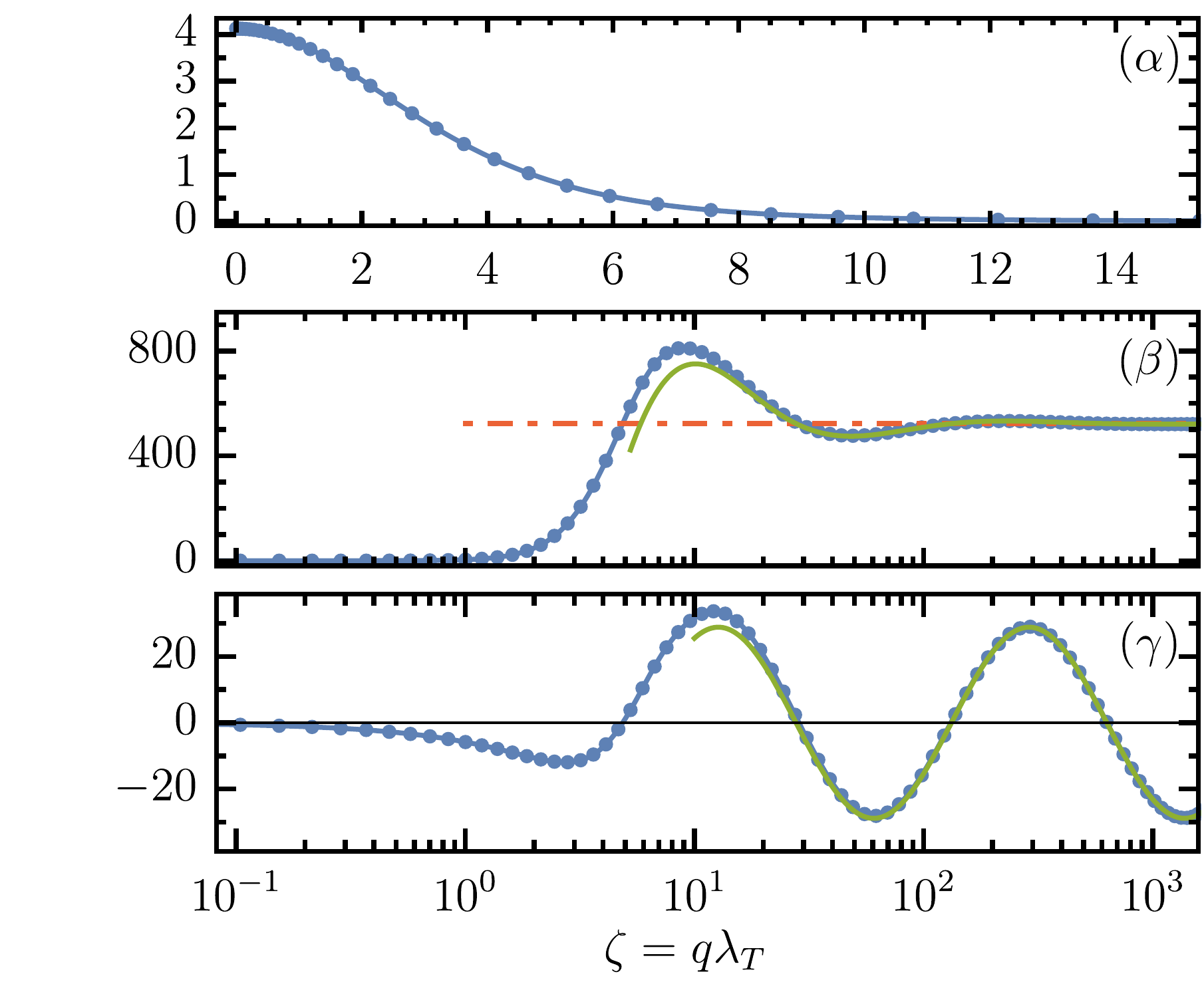}}}
\subfigure[\ $\kappa_*\lambda_T=5$, $\beta E = 3.98$]{\scalebox{0.32}{\includegraphics{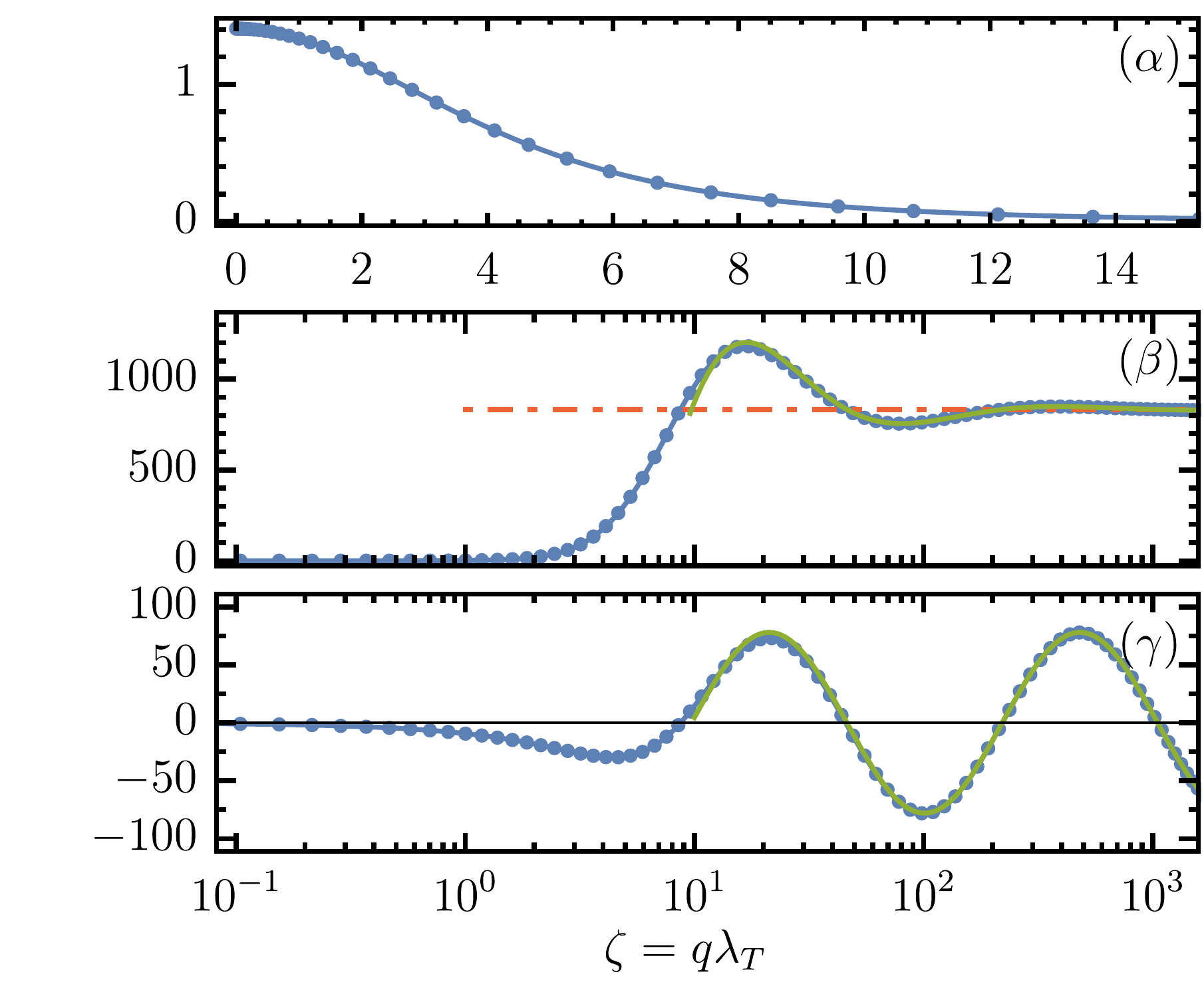}}}
\caption{(Color online) Full momentum distribution $n(\zeta)$ up to third order in the fugacity at unitarity with $\kappa_*\lambda_T=1,3$, and $5$. Fugacities have been chosen such that $z^3 e^{\beta E} = 0.6$. The notation is the same as in Fig.~\ref{fig:momentum}.}
\label{fig:momentumkappa}
\end{center}
\end{figure*}

\subsection{Two- and three-body contacts}\label{sec:contacts}

There exists a large set of universal relations that describe thermodynamic quantities and the short-time and short-distance structure of a quantum gas~\cite{tan08a,braaten11,Werner12}. The magnitude of these relations is set by the two-body contact $C_2$ and the three-body contact $C_3$, which are related to the energy of the system by the adiabatic theorems~\cite{tan08a,braaten11,Werner12}:
\begin{eqnarray}
C_2 &=& - 8 \pi m \frac{\partial E}{\partial a^{-1}} \biggr|_{\kappa_*}   =  \frac{V}{\lambda_T^4} \sum_{j\geq 2} c_{2,j} z^j  \\
C_3 &=& - \frac{m \kappa_*}{2} \frac{\partial E}{\partial \kappa_*} \biggr|_{a^{-1}}  =  \frac{V}{\lambda_T^5} \sum_{j\geq 3} c_{3,j} z^j ,
\end{eqnarray}
where $c_{2,j}=16\pi^2\partial b_j/\partial(\lambda_T/a)$ and $c_{3,j}=\pi\kappa_*\partial b_j/\partial \kappa_*$, and we abbreviate $c_{3,3}=c_3$. In Fig.~\ref{fig:contact}, we plot the intensive contact parameters $C_2/Nk_n$ [Fig.~\ref{fig:contact}(a)] and $C_3/Nk_n^2$ [Fig.~\ref{fig:contact}(b)] as a function of rescaled temperature $T/T_n$, where $k_n=(6\pi^2 n)^{1/3}$ and $T_n=k_n^2/2m$. For $T\gg T_n$, the contacts approach the asymptotic results $C_2/Nk_n=64T_n/3 T$ and $C_3/Nk_n^2=4s_0T_n^2/\sqrt{3}\pi^2T^2$ (green lines without dots in Fig.~\ref{fig:contact}]) As the temperature is lowered, both $C_2$ and $C_3$ strongly increase and then approach a constant value at low temperature. In addition, we show the three-body contact as a function of $\kappa_*\lambda_T$ in Fig.~\ref{fig:contact}(c). The numerical results agree with the analytical limits $c_3=3\sqrt{3}(\kappa_*\lambda_T)^2e^{\beta E_T}$ for large $\kappa_*\lambda_T$ and $c_3=3\sqrt{3}s_0$ for small $\kappa_*\lambda_T$~\cite{braaten11}.

\subsection{Momentum distribution}\label{sec:momentum}

A particular universal relation connects the two- and three-body contacts to the high-momentum tail of the momentum distribution~\cite{tan08b,Castin11,braaten11}:
\begin{equation}
n(q) = \frac{{\cal C}_2}{q^4} + \frac{{\cal C}_3}{q^5} F(q) + {\it O}(1/q^6), \label{eq:highmomentumtail}
\end{equation}
where ${\cal C}_{2/3}=C_{2/3}/V$ is the intensive contact density and $F(q)$ is a log-periodic function of the momentum given by $F(q)=A\sin(2s_0\ln q/\kappa_*+2\phi)$, where $A=89.26260$ and $\phi=-0.669064$. Indeed, a comparison of the experiment~\cite{makotyn14} with the asymptotic result~\eqref{eq:highmomentumtail} using a scaling ansatz for the contact parameters suggests that the observed momentum distribution is consistent with the three-body tail~\cite{hudson14}. Note that the momentum distribution can be computed for a {\it single} trimer~\cite{Castin11,Bellotti13}. Here, we provide a full calculation of the momentum distribution {\it at finite density} to third order in the fugacity.

The momentum distribution $n(q)$ is defined as the zero-time limit $G(0^-,q)$ of the imaginary-time propagator. Formally, the diagrams that contribute to $n(q)$ up to third order in $z$ are the same as for the density in Fig.~\ref{fig:diagrams}, although the black dot now denotes a momentum insertion $q$ into the diagram, changing the structure of the calculation fundamentally. Figure~\ref{fig:momentum} shows our result for the ${\it O}(z^3)$ part $n_3(q)$ of the momentum distribution at unitarity with $\kappa_*\lambda_T=3$, where we parametrize 
\begin{equation}
n(q)=e^{\beta E_T} \big[n_1(q)z+n_2(q)z^2+n_3(q)z^3+{\it O}(z^4)\big].
\end{equation}
Figure~\ref{fig:momentum}(a) shows $n_3(q)$ as a function of dimensionless momentum $\zeta = q\lambda_T$, Fig.~\ref{fig:momentum}(b) amplifies the high-momentum tail, and Fig.~\ref{fig:momentum}(c) shows the momentum distribution with a subtracted leading-order two-body term. Our results are in excellent agreement with the leading-order (red dashed line) and next-to-leading order (green line) Tan relations~\eqref{eq:highmomentumtail}, where the ${\it O}(z^3)$ contact parameters are extracted from the virial coefficients in an independent calculation, providing a strong check of our results. We remark that the detailed comparison of the numerical high-momentum tail with the independently obtained contact parameters over more than three orders of magnitude in momentum poses exceptional constraints on the numerical implementation, requiring a relative error of $10^{-9}$ in our simulations. Note that the onset of the asymptotic tail is set by $k_\kappa=\sqrt{2mE_T}$ rather than $k_n$, which marks the asymptotic 
regime in deeply 
degenerate Fermi gases~\cite{Stewart10}.

\begin{figure}[t!]
\begin{center}
\includegraphics[width=0.85\columnwidth]{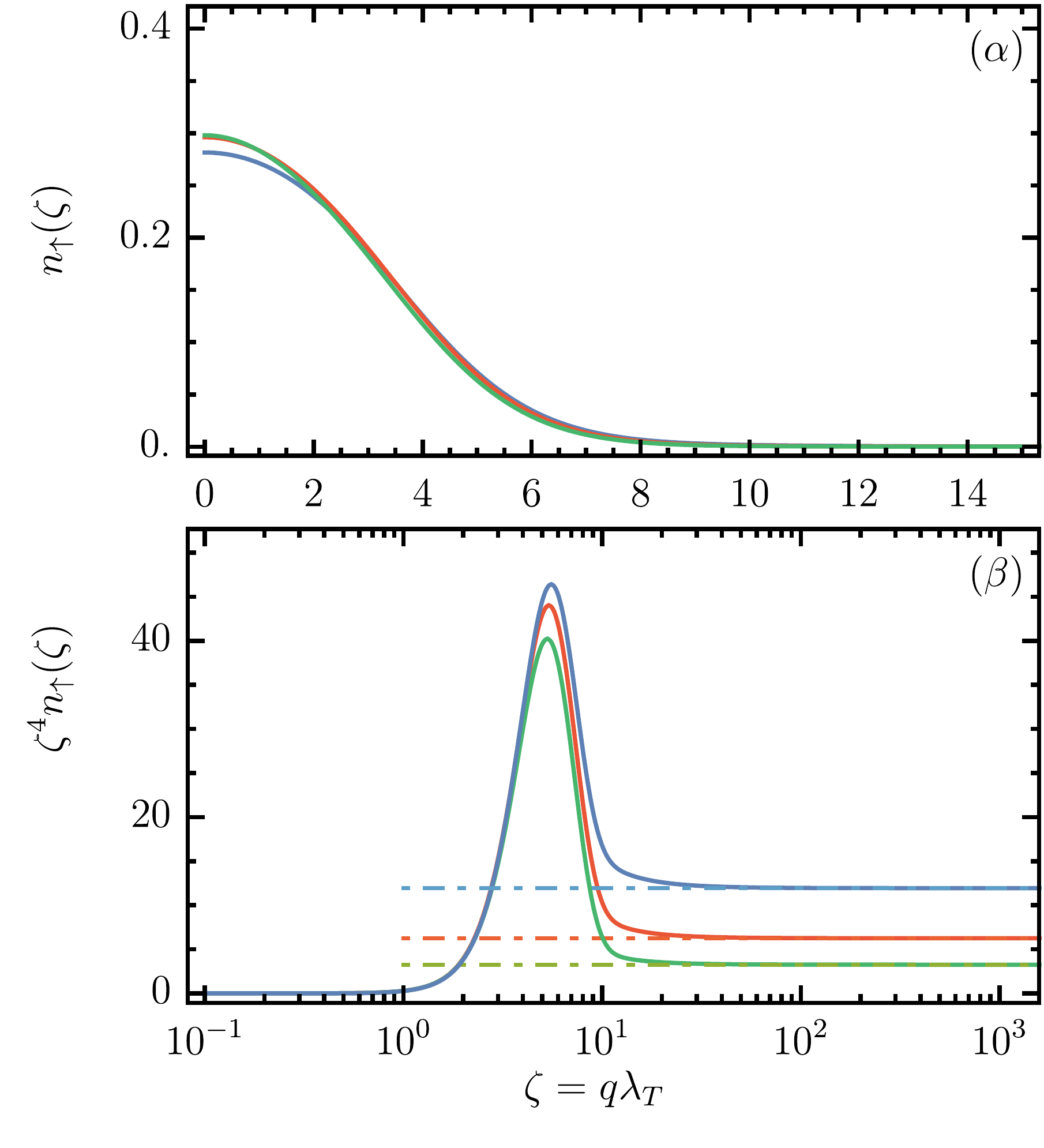}
\caption{(Color online) Momentum distribution of spin-up particles in a population-balanced two-component Fermi gas as a function of dimensionless momentum $\zeta = q\lambda_T$ including terms up to third order in the fugacity $z$. The scattering lengths are $\lambda_T / a = -1$ (green), $\lambda_T / a = 0$ (red) and $\lambda_T / a = 1$ (blue). Fugacities were chosen such that $z^2 e^{\beta E} = 0.15$, where $E = E_D$ for $a > 0$ and $E = 0$ otherwise. Dot-dashed lines denote the corresponding values of the contact as obtained from an independent calculation for the virial coefficients.}
\label{fig:momentumfermi}
\end{center}
\end{figure}

To illustrate the behavior of the momentum distribution as a function of scattering length and the three-body parameter, we show the full momentum distribution including all terms up to third order in the fugacity in Fig.~\ref{fig:momentumainv} for fixed $\kappa_*\lambda_T=3$ and three different scattering lengths $a/\lambda_T=-3.16,0$, and $3.16$. In Fig.~\ref{fig:momentumkappa}, the momentum distribution at unitarity is shown for three values of the three-body parameter, $\kappa_*\lambda_T = 1,3$, and $5$. In both Figs.~\ref{fig:momentumainv} and~\ref{fig:momentumkappa}, the fugacity was chosen such that $z^3 e^{\beta E} = 0.6$, where $E\geq0$ again denotes the deepest bound-state energy. All momentum distributions agree with the contact parameters at unitarity or finite scattering length as obtained from the adiabatic theorems using the virial coefficients, which were computed in an independent calculation. Figures~\ref{fig:momainv-a} and~\ref{fig:momainv-c} illustrate that, compared to the unitary case, 
the presence of another scale ($1/a \neq 0$) delays the saturation to the universal high-momentum behavior \eqref{eq:highmomentumtail} by almost an order of magnitude in momentum.

\begin{figure}[t!]
\begin{center}
\includegraphics[width=0.95\columnwidth]{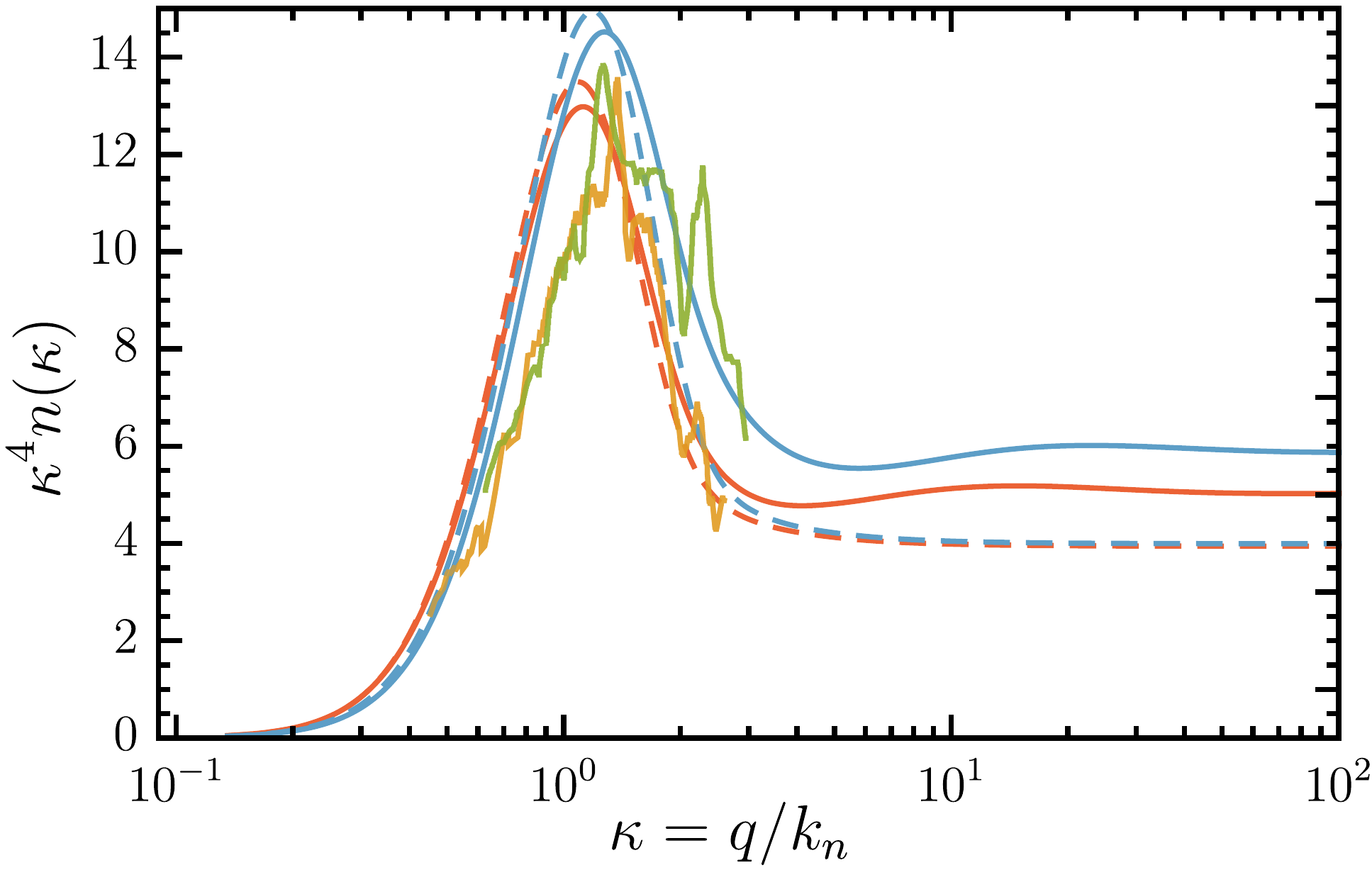}
\caption{(Color online) Momentum distribution of a unitary Bose gas. $\kappa=q/k_n$ denotes the dimensionless momentum. The solid red and blue lines are the result of the trap-averaged virial expansion with fugacities $z_1 = 0.5$ and $z_2=0.4$, respectively, corresponding to the experimental data~\cite{makotyn14} at $\langle n_1\rangle=5.5\cdot10^{12}\,{\rm cm}^{-3}$ (orange line) and $\langle n_2 \rangle=1.6\cdot10^{12}\,{\rm cm}^{-3}$ (green line). The dashed red and blue lines are the corresponding ${\it O}(z^2)$ results. All momentum distributions are normalized to unity $\int d^3\kappa\,n(\kappa)/(2\pi)^3=1$.}
\label{fig:momentumexp}
\end{center}
\end{figure}

To conclude this section, we briefly make a comparison with the fermionic case, which can be computed along the lines outlined in the appendices. For an approach that calculates the momentum distribution via the spectral function, see~\cite{sun15}. Figure~\ref{fig:momentumfermi} shows the momentum distribution $n_\uparrow(\zeta)$ of spin-up particles in a two-component Fermi gas with equal population in both spin species, calculated to third order in the fugacity. The three different scattering lengths are $\lambda_T / a = -1,0,1$, and the fugacities were chosen such that $z^2 e^{\beta E} = 0.15$. While at first glance the momentum distributions look qualitatively the same as the bosonic ones, the fermionic system shows no oscillatory behavior in its high-momentum tail, as is to be expected since the three-body contact is without analog in fermionic gases with equal masses.

\section{Comparison with experiments}\label{sec:experiment}

We compare our results with the experiment by Makotyn {\it et al.}~\cite{makotyn14}, which measures the momentum distribution following a quench to the unitary limit for two different initial densities $\langle n_1\rangle=5.5\cdot10^{12}\,{\rm cm}^{-3}$ and $\langle n_2\rangle=1.6\cdot10^{12}\,{\rm cm}^{-3}$. Following Ref.~\cite{laurent14}, we assume a constant phase space density $n \lambda_T^3$ and average our results over a Thomas-Fermi density profile. Since our calculations are performed at fixed $\kappa_*\lambda_T$, we keep $\kappa_*/k_n$ fixed at its value at the trap center, which neglects logarithmic corrections to the fugacity. The experiment was performed using ${}^{85}$Rb, which has a three-body parameter $\kappa_*=38(1)\,\mu m^{-1}$~\cite{wild12}. Remarkably, our results are in good agreement with the experiment if and only if we exclude the lowest trimer branch from our calculation and set $\kappa_*'=\kappa_*/22.7$, indicating that on the time scales of the experiment~\cite{makotyn14}, the 
lowest Efimov trimer branch is not occupied. In this case, the virial expansion agrees well with the experimental data with $z_1=0.5$ and $z_2=0.4$. The small values of the fugacity justify the use of the virial expansion. The results are shown in Fig.~\ref{fig:momentumexp}. For comparison, we include a fit up to second order in $z$ as a dashed line.

\section{Results and Conclusions}\label{sec:summary}

In summary, we have characterized the strongly interacting Bose gas in the normal phase by computing the first three virial coefficients, the two-body and the three-body contacts, and the momentum distribution. The results for the momentum distribution are in good quantitative agreement with the recent experiment~\cite{makotyn14}. Surprisingly, the fit indicates that the lowest Efimov trimer state is not populated in the experiment. Our work opens the possibility for the spectroscopy of Efimov states at unitarity.
\begin{acknowledgments}
We thank Christian Langmack and Wilhelm Zwerger for discussions. This work is supported by the DFG research unit "Strong Correlations in Multi-flavor Ultracold Quantum Gases"  (M.B.) and by LPS-MPO-CMTC, JQI-NSF-PFC, and ARO-MURI (J.H.). The authors acknowledge the University of Maryland supercomputing resources (www.it.umd.edu/hpcc) made available in conducting the research reported in this paper.
\end{acknowledgments}

\appendix
\section{STM equation}\label{sec:appA}

The ${\it O}(z^3)$ of the virial expansion relates the density and the momentum distribution to the vacuum three-body $T_3$ matrix, and we summarize some basic properties in this section. The $T_3$ matrix is obtained by solving the Skorniakov-Ter-Martirosian (STM) integral equation \cite{skorniakov57}, which is shown diagrammatically in Fig. 1(a). The three-body matrix depends on the total energy $s$ and the total momentum ${\bf P}$ of the incident atom-dimer state as well as the energy and momentum $(E_{\rm in}, {\bf k}_{\rm in})$ and $(E_{\rm out}, {\bf k}_{\rm out})$ of the ingoing and outgoing atom line. It turns out that our calculations require only the on-shell $T_3$ matrix where the external atom energy is equal to the kinetic energy of a free atom, i.e., $E_{\rm in} = \varepsilon_{\bf k} \equiv k^2/2m$ and $E_{\rm out} = \varepsilon_{\bf p} \equiv p^2/2m$. Using the Feynman rules of the Lagrangian~(2) in the main text, the integral equation for the on-shell $T_3$ matrix with total energy $s$ and 
incoming and outgoing atom momenta $\vp$ and $\vk$ reads
\begin{align}
 T_3 (s,& \, \eps_\vp,\eps_\vk, \vP,\vp,\vk)   = \left[ \frac{1}{s - \eps_\vp - \eps_\vk - \eps_{\vP-\vp-\vk}} + \frac{g_3}{9g_2^2} \right] \nonumber \\
 & + \int_\vq  \left[ \frac{1}{s - \eps_\vp - \eps_\vq - \eps_{\vP-\vp-\vq}} + \frac{g_3}{9g_2^2} \right] \nonumber \\
 &\quad \times T_2 \left(s- \eps_\vq - \frac{\eps_{\vP -\vq}}{2} \right) T_3 (s,\eps_\vq,\eps_\vk, \vP,\vq,\vk)
\label{eq:3body-STM-equation-with-cms-P} ,
\end{align}
where we have defined $\int_\vk \equiv \int d^3k / (2\pi)^3$ as a shorthand for the momentum integrals. $T_2$ is the two-particle scattering matrix in vacuum:
\begin{align}
 T_2(s) & = \frac{8\pi}{m} \frac{1}{\frac{1}{a} - \sqrt{-m s}} . \label{eq:3body-T2matrix-bosons-final}
\end{align}
Due to the Galilean invariance of the vacuum state, we have the following important transformation property:
\begin{align}
  T_3 (s, \eps_\vp,\eps_\vk, \vP,\vp,\vk) = t_3 \left(s - \frac{\eps_\vP}{3},\vp - \frac{\vP}{3}, \vk - \frac{\vP}{3} \right) , \label{eq:3body-T3-Gallilean-invariance-trafo}
\end{align}
where $t_3 (s,\vq,\vq') = T_3 (s, \eps_\vq, \eps_{\vq'}, {\bf 0}, \vq,\vq' )$ denotes the $T_3$ matrix in the center-of-mass frame. The STM equation can be decoupled in angular momentum channels using the angular decomposition
\begin{align}
 t_3 (s,\vp,\vk) & =  \sum_l (2l+1) P_l (\cos \theta) t_3^{(l)} (s,p,k) \label{eq:3body-T3-intermsof-angularT3} \\
 t_3^{(l)} (s,p,k) & =  \frac{1}{2} \int_{-1}^1 d\cos \theta \, P_l (\cos \theta) t_3 (s,\vp,\vk) , \label{eq:3body-T3-angular-projection}
\end{align}
where $P_l (\cos \theta)$ are the Legendre polynomials and $\cos \theta = \hat{\vp} \cdot \hat{\vk}$, where the hat denotes a unit vector. This yields a set of decoupled STM equations for the different angular momentum channels:
\begin{align}
& t_3^{(l)} (s,p,k)  = \left[ \frac{m}{pk} Q_l\left( \frac{m}{pk} \left[ s - \frac{p^2}{m} - \frac{k^2}{m} \right] \right) - m \frac{H (\Lambda)}{\Lambda^2} \delta_{l0} \right] \nonumber \\
                   &  \quad + \frac{4}{\pi m} \int_0^\Lambda dq \frac{q^2}{\frac{1}{a} - \sqrt{-m s + \frac{3}{4} q^2}}  t_3^{(l)} (s,q,k) \nonumber \\
& \quad            \times        \left[ \frac{m}{pq} Q_l\left( \frac{m}{pq} \left[ s - \frac{p^2}{m} - \frac{q^2}{m} \right] \right) - m \frac{H (\Lambda)}{\Lambda^2} \delta_{l0} \right] .
 \label{eq:3body-STM-eq-bosons-angularT3}
\end{align}
Here, $Q_l(z)$ denotes a Legendre function of the second kind \cite{oberhettinger_1966}. As discussed in the main text  (see also~\cite{braaten06}), we renormalize the theory in order to eliminate the bare coupling constants in favor of the log-periodic cutoff function $H(\Lambda)$. The numerical solution of the integral equation \eqref{eq:3body-STM-eq-bosons-angularT3} in the $l=0$ channel yields an Efimov spectrum with infinitely many three-body bound states and a discrete scaling symmetry as described in the Introduction of the main text.

\section{The momentum distribution}\label{sec:appB}

\subsection{Expansion of the Green's function}
In the following, we will illustrate the calculation for the momentum distribution $n(\vq)$.
The momentum distribution is defined as the zero-time limit of the imaginary-time Green's function:
\begin{equation}
 n (\vq) = -\lim_{\tau \rightarrow 0^-} G(\tau, \vq) .\label{eq:momdis-greensfct-relation}
\end{equation}
The total density is defined as
\begin{equation}
 n = \int_{\bf q} n({\bf q}) = -\lim_{\tau \rightarrow 0^-} \int_{\bf q} G(\tau, \vq), \label{eq:density-greensfct-relation}
\end{equation}
which implies that the results of the diagrams for the density can be obtained via integration of the results for the momentum distribution over all momenta.
The noninteracting Green's function $G_0$ can be expanded in the fugacity by expanding the Bose-Einstein distribution in powers of $z$:
\begin{align}
 G_0(\tau,\vq) & =  - e^{-\tau (\eps_\vq - \mu)} \bigg[ \Theta (\tau) + n_B (\eps_\vq -\mu)\bigg] \nonumber \\
& = \sum_{j=0}^{\infty} z^j G^{(j)}_0 (\tau,\vq) .\label{eq:3body-bosepropagator-expanded}
\end{align}
Diagrammatically, the ${\it O}(z^n)$ contribution $G^{(j)}_0 (\tau,\vq)$ in the expansion of the free Green's function is denoted by a line that is slashed $n$ times. Since the zero-order contribution is purely retarded, all backward-running lines contribute at least one power of $z$ to a diagram. The terms that define the right-hand side of Eq. \eqref{eq:3body-bosepropagator-expanded} up to third order in the fugacity yield the one-body diagrams shown in Figs. 1(b)-1(d) of the main text. For the momentum distribution $n(\vq)$, the external lines have a momentum $\vq$ which is not integrated over.

Like for fermions~\cite{leyronas11}, the $j$ th order in the virial expansion of a correlator can be constructed systematically by constructing the Feynman diagrams with a maximum number of $j$ backward-propagating lines. These diagrams involve an infinite number of repeated scattering of two or three forward-propagating lines, which, when resummed, yield the vacuum two-body $T_2$ and three-body $T_3$ matrices. The virial expansion of the density gives the virial coefficients and, using the adiabatic theorems, the two- and three-body contact parameters as discussed in the main text. Our calculation, which is set up for the full Green's function, allows us to go beyond these results to compute the full momentum distribution, which contains universal Efimov correlations. In the following, we discuss the derivation of the diagrammatic contributions to the momentum distribution up to third order in $z$ and state the results for all diagrams.

\subsection{One-body diagrams}
The contribution to a one-body diagram that is slashed $\ell$ times can be directly inferred from Eq.~\eqref{eq:3body-bosepropagator-expanded}:
\begin{eqnarray}
  \lim_{\tau \rightarrow 0^-} G^{(\ell)}_0 (\tau,\vq) =  - e^{- \ell \beta \eps_\vq}, \label{eq:3body-1p-bosons-momdis}
\end{eqnarray}
where Fig. 1(b) has $\ell = 1$, Fig. 1(c) corresponds to $\ell = 2$, and Fig. 1(d) corresponds to $\ell = 3$.

\subsection{Two-body diagrams}
We will consider the ${\it O}(z^2)$ two-body diagram in some detail, and quote the results for the remaining diagrams for completeness. We denote the imaginary-time difference between time zero and the scattering vertex by $t_1$, while the scattering vertex shall act over a time difference $t_2$. With these conventions, Fig. 1(e) reads:
\begin{align}
 \raisebox{-0.4cm}{{\scalebox{0.4} {\includegraphics{n2p2.pdf}}} } & =  - z^2 \int_A dt_1 dt_2 \int_\vk G^{(0)}_0 (t_1,\vq) G^{(1)}_0 (-t_2,\vk) \nonumber \\
 &\quad  \times  e^{2\mu t_2} T_2(t_2,\vq+\vk)  G^{(1)}_0 (-t_1-t_2,\vq) \nonumber \\
     & =  z^2 \int_A dt_1 dt_2 \int_\vq T_2 (t_2,\vq+\vk) e^{-t_1 (\eps_\vq + \eps_\vk)} \nonumber \\
     &\quad \times e^{-(\beta-t_1-t_2) (\eps_\vq + \eps_\vk)}, \label{eq:3body-2a-time-form}
\end{align}
where $A = \lbrace (t_1,t_2): 0 < t_1+t_2 < \beta, 0 < t_1, 0 < t_1 \rbrace$.
Using the generalized convolution theorem for Laplace transforms \cite{leyronas11,doetsch_2012}, we can write this as the inverse Laplace transform of the Laplace transforms of the three functions in the integrand:
\begin{align}
 \raisebox{-0.4cm}{{\scalebox{0.4} {\includegraphics{n2p2.pdf}}} } & =  z^2 \int_s e^{-\beta s} \int_\vk \frac{T_2 (s,\vk+\vq)}{(s-\eps_\vk-\eps_\vq)^2} .
 \label{eq:3body-diagram2a-symbolic-step}
\end{align}
The integration $\int_s = \int_{BW} ds / 2 \pi i$ denotes a Bromwich contour in the complex energy plane, which runs (parallel to the imaginary axis) to the left of the dimer pole and the branch cut of the $T_2$ matrix. We now substitute $\vp = \vq+\vk$ and $s \rightarrow s - \eps_\vp$, which allows us to eliminate the angle integrations from the momentum integration and to transform the above expression into
\begin{align}
  &\raisebox{-0.4cm}{{\scalebox{0.4} {\includegraphics{n2p2.pdf}}} } =  z^2 \int_s e^{-\beta s} T_2(s) \int_\vp \frac{e^{-\frac{\beta}{2} \eps_\vp}}{(s + \frac{\eps_\vp}{2} - \eps_{\vq-\vp} - \eps_\vq)^2} \nonumber \\
 & =  \frac{z^2}{(2\pi)^2} \int_s  \int_0^\infty dp \, \frac{2 e^{-\beta s} T_2(s) p^2 e^{-\beta \frac{p^2}{4m}}}{\left(\frac{p^2}{4m}\right)^2 + \left(\frac{q^2}{m} - s\right)^2 - \frac{p^2}{2m} \left( \frac{q^2}{m} +s \right)}. \nonumber \\ \label{eq:3body-diagram2a-final}
\end{align}
The remaining two-body diagrams can be evaluated in a similar way, and we list only the results:
\begin{align}
 \raisebox{-0.4cm}{{\scalebox{0.4} {\includegraphics{n2p3a.pdf}}} } & = \raisebox{-0.4cm}{{\scalebox{0.4} {\includegraphics{n2p3c.pdf}}} } =  z e^{-\beta \eps_\vq} \raisebox{-0.4cm}{{\scalebox{0.4} {\includegraphics{n2p2.pdf}}} } \label{eq:3body-momdis-bosons-diagram-2d}
\end{align}
\begin{align}
 &\qquad\qquad\qquad\qquad\qquad\raisebox{-0.4cm}{{\scalebox{0.4} {\includegraphics{n2p3b.pdf}}} } \nonumber \\
 & =  \frac{z^3}{2\pi^2} \int_s  \int_0^\infty dp \frac{p^2 e^{-\beta (s + \frac{3p^2}{4m} + \frac{q^2}{6m})} T_2(s)}{\left( s-\frac{p^2}{4m} \right)^2 - \frac{8}{9} \frac{q^2}{m} \left( s + \frac{p^2}{4m} \right) + \frac{16}{81} \frac{q^4}{m^2} } .
 \label{eq:3body-momdis-bosons-diagram-2c}
\end{align}

\subsection{Three-body diagrams}
The three-body diagrams differ in various ways. First and foremost, note that Fig. 1(i) gets an overall symmetry factor of $1/2$, while Figs. 1(j)-1(m) share the same factor of $1$. In addition, the lack of the integration over the external momentum considerably complicates the evaluation of the diagrams both numerically and analytically compared to the density. Eliminating the convolution integral over the imaginary times in favor of a single complex integration, we find for Fig. 1(i)
\begin{align}
   &\raisebox{-0.8cm}{{\scalebox{0.25} {\includegraphics{n3p1.pdf}}} }  =   \frac{z^3}{2} \int_s e^{-\beta s} \int_{\vk,\vp} T_2^2 (s-\eps_\vq,\vk+\vp) \nonumber \\
   & \quad \quad \quad \quad \quad \quad \times\frac{T_3(s,\eps_\vq,\eps_\vq,\vk+\vp+\vq,\vq,\vq)}{(s-\eps_\vk-\eps_\vp-\eps_\vq)^2},
            \label{eq:3body-bosons-diagram3a-raw}
\end{align}
where $\vk$ and $\vp$ denote the loop momenta of the two backward-running atom propagators that are connected to the dimer fields. In the case of the three-particle diagrams, the Bromwich contour runs to the left of the trimer pole which we have chosen to be the physically deepest one. This amounts to an implicit subtraction of all deeper trimer poles. We now substitute
\begin{eqnarray}
 s' & = & s -\frac{\eps_{\vq+\vp+\vk}}{3} \label{eq:3body-3psector-s-subst} \\
\left( \begin{array}{l}
\vp'\\
\vk'
\end{array}
\right)
 & = &
\left(
\begin{array}{ll}
 \vspace{2pt} -\frac{1}{3}  \mathbbm{1}_3 & -\frac{1}{3} \mathbbm{1}_3 \\
-\frac{1}{2} \mathbbm{1}_3 & \frac{1}{2} \mathbbm{1}_3
\end{array}
\right)
\left(
\begin{array}{l}
\vp-\vq\\
\vk-\vq
\end{array}
\right) \nonumber \\
&=& 
\left(
\begin{array}{l}
\vspace{2pt} \frac{2}{3} \vq - \frac{1}{3} \vk - \frac{1}{3} \vp\\
\frac{1}{2} \vk -\frac{1}{2}\vp
\end{array}
\right),
\end{eqnarray}
where the determinant of the transformation contributes a factor $3^3$ and $\mathbbm{1}_3$ denotes the identity matrix in three dimensions. Using Eq.~\eqref{eq:3body-T3-Gallilean-invariance-trafo} and relabeling $s' \rightarrow s, \vk' \rightarrow \vk, \vp' \rightarrow \vp$, the contribution of Fig.~1(i) can be expressed in terms of the center-of-mass amplitudes as
\begin{widetext}
\begin{align}
 \raisebox{-0.8cm}{ {\scalebox{0.25} {\includegraphics{n3p1.pdf}}} }   = & \frac{3^3 z^3}{2} \int_s e^{-\beta s} \int_{\vk,\vp} e^{-3\beta \eps_{\vq-\vp}} \frac{T_2^2 \left( s - \frac{3}{2} \eps_\vp \right) t_3 (s,\vp,\vp)}{(s- \frac{3}{2} \eps_\vp - 2\eps_\vk)^2} \nonumber \\
  = & \frac{3^2 z^3 m^3}{4 q \beta (2\pi)^3} \sum_l (2l+1) \int_s e^{-\beta s} \int_0^\infty dp p \frac{T_2^2\left( s - \frac{3p^2}{4m} \right) t_3^{(l)}(s,p,p)}{\sqrt{-ms + \frac{3p^2}{4}}}e^{-\frac{3\beta}{2m} (q^2 + p^2)} \sinh\left( \frac{3pq\beta}{m} \right) \label{eq:3body-momdis-diagram3a-final}.
\end{align}
For the last line, we have solved the elementary integral over $\vk$ and used the fact that $t_3(s,\vp,\vp) = \sum_l P_l (\hat{\vp} \cdot \hat{\vp}) t_3^{(l)} (s,p,p)$ contains only forward scattering, such that all the Legendre polynomials equal to $1$. We have then integrated over the angles of $\vp$. We will now list the resulting expressions for the other diagrams, which are obtained in an analogous fashion:
\begin{align}
 \raisebox{-0.8cm}{{\scalebox{0.25} {\includegraphics{n3p2.pdf}}} }  = & \frac{3^2 z^3 m}{(2\pi)^4 \beta} \sum_l (2l+1) \int_s e^{-\beta s} \int_0^\infty dp \int_0^\infty dk \, p k  \,  t_3^{(l)}(s,k,k)  \frac{ \left[ e^{-\frac{3\beta}{8m} (k-p)^2} - e^{-\frac{3\beta}{8m} (p+k)^2} \right] \, T_2^2\left( s - \frac{3}{4m} k^2 \right)}{\left[ \frac{3k^2}{4m} + \frac{(p-2q)^2}{4m} - s \right] \left[ \frac{3k^2}{4m} + \frac{(p+2q)^2}{4m} - s \right]}, \label{eq:3body-momdis-diagram3b-final}
\end{align} 
\begin{align}
 \raisebox{-0.8cm}{{\scalebox{0.25} {\includegraphics{n3p3.pdf}}} }   = & \frac{2 m^3}{\beta q} \frac{3^2 z^3}{(2\pi)^4} \sum_{l} (2l+1) \int_s e^{-\beta s} \int_0^\infty \frac{dp}{p} \int_0^\infty dk \, T_2 \bigg(s-\frac{3}{4m}k^2\bigg) T_2 \bigg(s-\frac{3}{4m}p^2\bigg) \nonumber \\
                     & \hspace{3cm} \times\bigg[ e^{-\frac{3\beta}{2m} (p-q)^2 } - e^{-\frac{3\beta}{2m} (p+q)^2 } \bigg] \tilde{Q}_l \bigg( \frac{m}{pk} \left[s - \frac{p^2}{m} - \frac{k^2}{m}\right] \bigg) t_3^{(l)} (s,p,k), \label{eq:3body-momdis-diagram3c-final}
\end{align}
where $\tilde{Q}_l(z) = - d Q_l (z) / dz$ is the solution of the angle integration. Furthermore, we have:
\begin{align}
  \raisebox{-0.8cm}{{\scalebox{0.25} {\includegraphics{n3p4.pdf}}} } 
  = & \frac{2 m^3}{\beta q} \frac{3^2 z^3}{(2\pi)^4} \sum_{l} (2l+1) \int_s e^{-\beta s} \int_0^\infty \frac{dp}{p} \int_0^\infty dk \, T_2 \bigg(s-\frac{3}{4m}k^2\bigg) T_2 \bigg(s-\frac{3}{4m}p^2\bigg)  \nonumber \\
                     & \hspace{3cm} \times\bigg[ e^{-\frac{3\beta}{2m} (p-q)^2 } - e^{-\frac{3\beta}{2m} (p+q)^2 } \bigg] \tilde{Q}_l \bigg( \frac{m}{pk} \left[s - \frac{p^2}{m} - \frac{k^2}{m}\right] \bigg) t_3^{(l)} (s,k,p) . 
 \label{eq:3daftersubstitutions}
\end{align}
For the last diagram, the integration over the angles can only be partially performed analytically:
\begin{align}
    \raisebox{-0.8cm}{{\scalebox{0.25} {\includegraphics{n3p5.pdf}}} }  = & \frac{3^3 z^3}{(2\pi)^4} \frac{2m}{3\beta q} \int \frac{ds}{2\pi i} e^{-\beta s} \int_0^{\infty} dk \, k^2 \int_0^{\infty} dp\, p^2 \int_{-1}^1 d\cos\theta  \nonumber \\
		     	 &  \hspace{3cm} \times e^{- \frac{3 \beta}{2m} (k^2 + p^2 + q^2)}  T_2 (s - \frac{3}{4m}k^2) T_2 (s - \frac{3}{4m} p^2) \sum_l (2l+1) t_3^{(l)} ( s, k, p) \nonumber \\
		     	 &  \hspace{3cm} \times\frac{   P_l( \cos\theta ) e^{-\frac{3\beta}{m} k p \cos\theta} }{(s - \frac{k^2}{m} -\frac{p^2}{m} - \frac{k p}{m} \cos\theta )^2} \frac{\sinh\bigg( \frac{3\beta q}{m} \sqrt{k^2+p^2+2pk\cos\theta} \bigg)}{\sqrt{k^2+p^2+2pk\cos\theta}} . \label{eq:3body-momdis-diagram3e-final}
\end{align}
\end{widetext}

\subsection{Numerical evaluation}
During the numerical evaluation, we have retained angular momenta up to $l = 10$, finding no change when including higher harmonics into the calculation. In each angular momentum channel, the STM equation \eqref{eq:3body-STM-eq-bosons-angularT3} is solved at different complex energies that lie on the Bromwich contour. Afterwards, all Feynman diagrams discussed in the previous sections are integrated numerically over the loop momenta and the complex energies using the numerical solution of the STM equations.

\bibliography{bib}

\end{document}